\documentclass[aps,prl,twocolumn,amsmath,amssymb]{revtex4-1}
\usepackage{amssymb}
\usepackage{bm}
\usepackage{url}
\usepackage{graphicx}
\usepackage{color}

\begin{document}

\title{Suppression of Rayleigh-Taylor turbulence by time-periodic acceleration}

\author{G. Boffetta}
\affiliation{Dipartimento di Fisica and INFN, Universit\`a di Torino, 
via P. Giuria 1, 10125 Torino, Italy}
\author{M. Magnani}
\affiliation{Dipartimento di Fisica, Universit\`a di Torino, 
via P. Giuria 1, 10125 Torino, Italy}
\author{S. Musacchio}
\affiliation{Universit\'e C\^ote d'Azur, CNRS, LJAD, Nice 06108, France}

\begin{abstract}
The dynamics of Rayleigh-Taylor turbulence convection in presence 
of an alternating, time periodic acceleration is studied by means
of extensive direct numerical simulations of the Boussinesq equations.
Within this framework, we discover a new mechanism of relaminarization
of turbulence:
The alternating acceleration, which initially produces a growing turbulent mixing layer,
at longer times suppresses turbulent fluctuation 
and drives the system toward an asymptotic stationary configuration. 
Dimensional arguments and linear stability theory are used to 
predicts the width of the mixing layer in the asymptotic state 
as a function of the period of the acceleration.
Our results provide an example of simple control and suppression of 
turbulent convection with potential applications in different fields.
\end{abstract}

%\pacs{,,}

\maketitle 

%%%%%%%%%%%%%%%%%%%%%%%%%%%%%%%%%%%%%%%%%%%%%%%%%%%%%%%%%%%%%%%%%
The acceleration of a mass of fluid against another one with lower density 
gives rise to the Rayleigh-Taylor (RT) instability at the interface
between the two 
fluids
\cite{chandrasekhar2013hydrodynamic,sharp1984overview,abarzhi2010review}. 
If the acceleration persists, the unstable interface evolves into a turbulent
mixing layer which broadens in time. 
RT turbulence is one of the paradigmatic examples of turbulent
mixing \cite{chertkov2003phenomenology,
boffetta2017incompressible,zhou2017rayleigh}
with a large variety of
applications, ranging from astrophysical and geophysical phenomena
(e.g. supernovae flames~\cite{hillebrandt2000type,bell2004direct}, 
Earth's mantel motions~\cite{neil1999rayleigh}. 
ionosphere irregularities~\cite{sultan1996linear})
to confined nuclear
fusion~\cite{kilkenny1994review,tabak1994ignition}, 
plasma physics~\cite{tabak1994ignition}, 
and laser matter interactions~\cite{taylor1996measurement}. 

Although in many instances the driving acceleration may be considered
constant (as in the case of gravity), there are important applications where
acceleration changes in time and may even reverse sign. Examples include
Inertial Confined Fusion \cite{nakai1996principles,betti2007shock}, 
pulsating stars and 
supernovae explosions \cite{colgate1966hydrodynamic}.
RT instability and turbulence with 
complex acceleration history have been studied experimentally 
\cite{dimonte2007rayleigh} and numerically 
\cite{livescu2011direct,ramaprabhu2016evolution,aslangil2016numerical}.
In particular, the effects of a single stabilizing phase (corresponding to 
gravity reversal) has been shown to produce a slowing down of the 
turbulent mixing layer \cite{dimonte2007rayleigh}
and an increase of small scale mixing \cite{ramaprabhu2013rayleigh}.

In this Letter we address the question of what happens to 
RT turbulence in presence of a time periodic acceleration
which continuously alternates phases of unstable and stable stratification.
By extensive numerical simulations of the Boussinesq model,
we find that the turbulent mixing layer, which initially
develops and grows quadratically in time, {\em eventually stops for any 
period of the acceleration modulation}. 
The mechanism at the basis of this surprising suppression of RT mixing
is the decay of the turbulent fluctuations during the stable
phase, with a reduction of the temperature/density fluctuations in the mixing
layer. 
We find simple scaling laws for the arresting time and the asymptotic
size of the mixing layer, and we show how the suppression of turbulence
can be understood in terms of linear stability theory.

%%%%%%%%%%%%%%%%%%%%%%
\begin{figure}[h]
\includegraphics[width=0.3\columnwidth]{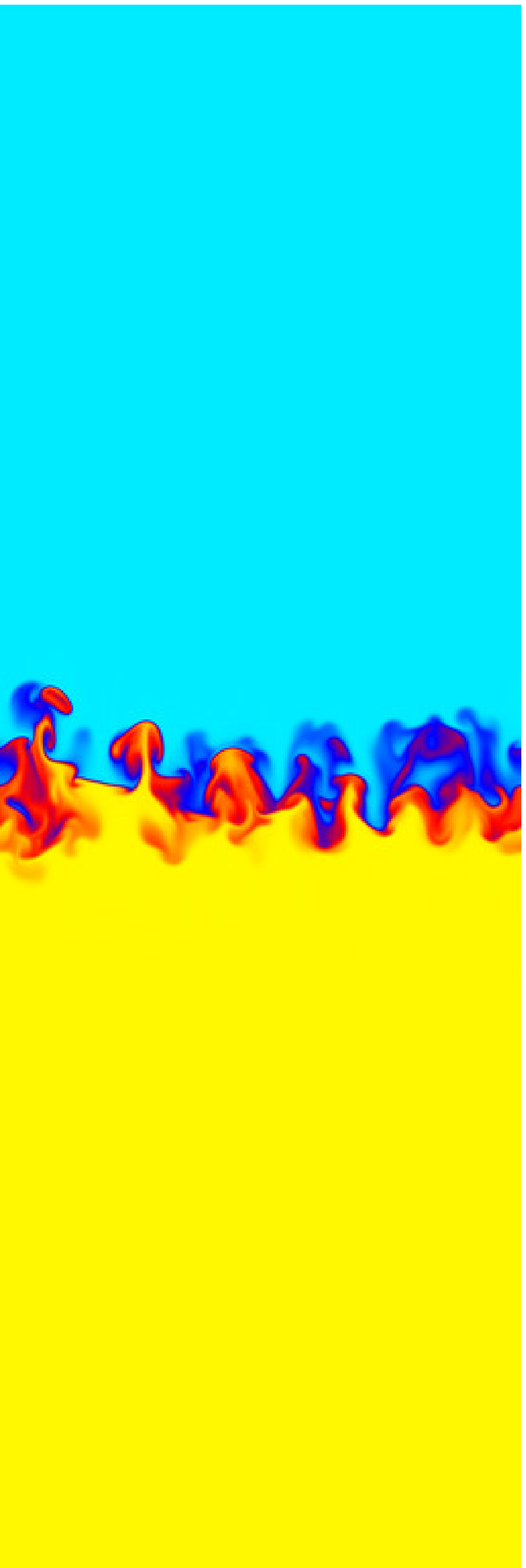}
\includegraphics[width=0.3\columnwidth]{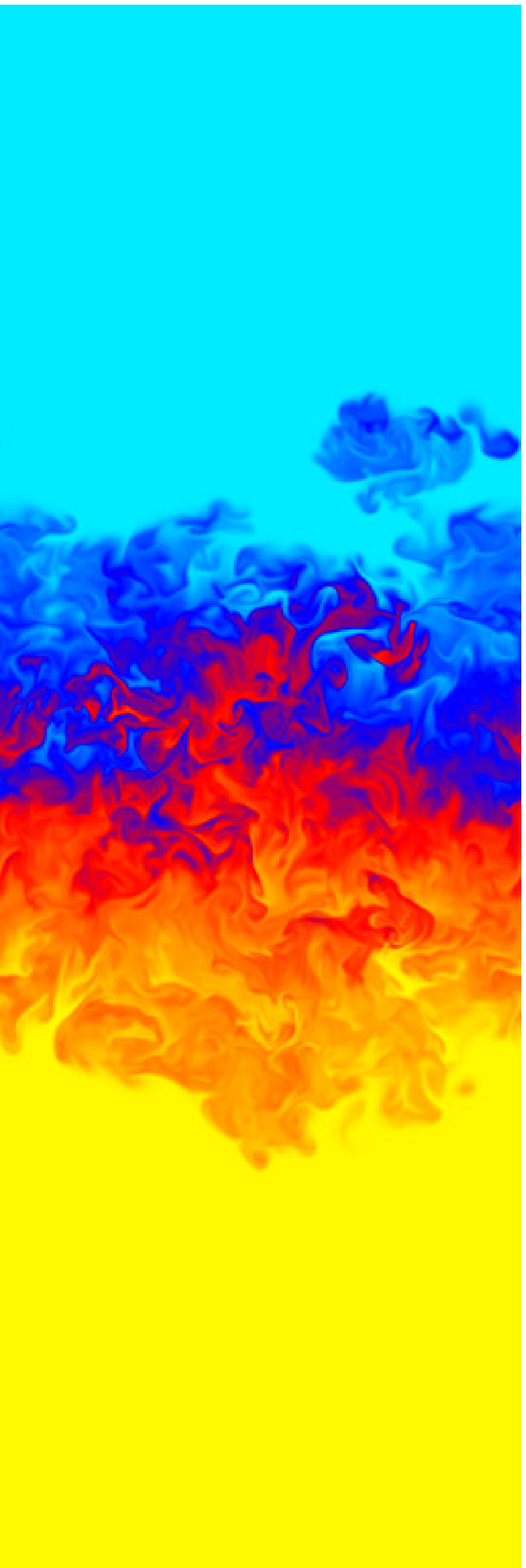}
\includegraphics[width=0.3\columnwidth]{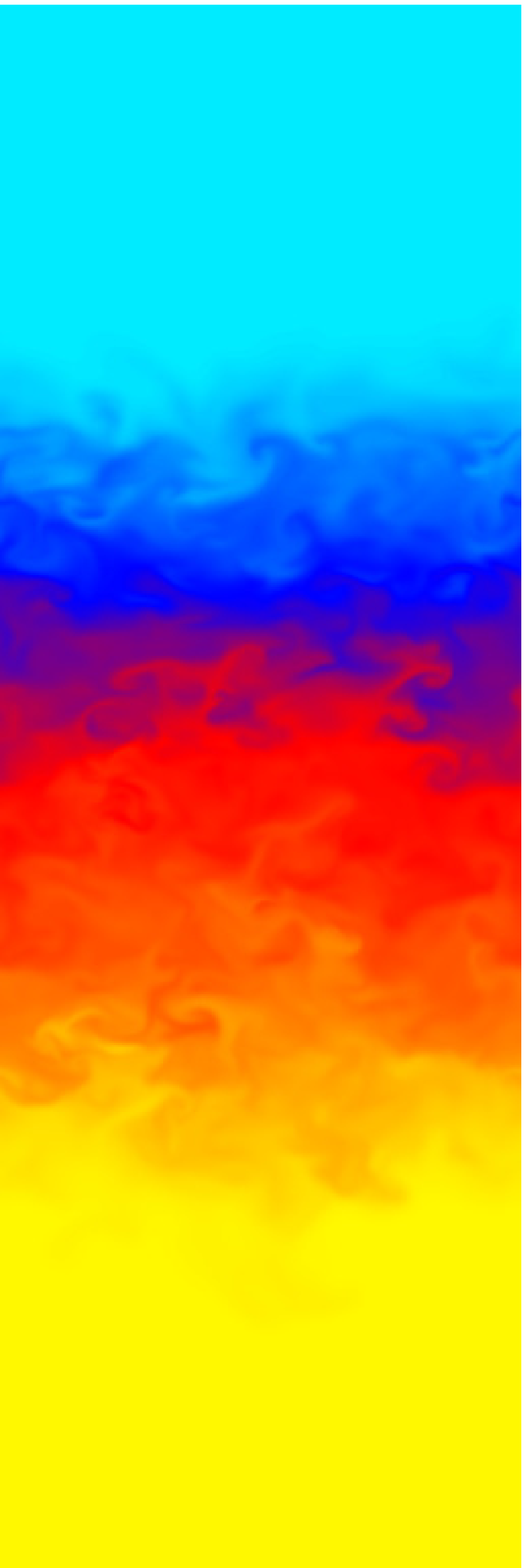}
\caption{Vertical sections of the temperature field (yellow hot, 
blue cold) for 
Rayleigh-Taylor turbulence in periodic acceleration of period $T=3 \tau$ at
times $t=1.5 T$ (left), $t=4 T$ (center) and $t=10 T$ (right).}
\label{fig1}
\end{figure}
%%%%%%%%%%%%%%%%%%%%%%

%%%%%%%%%%%%%%%%%%%%%%%%%%%%%%%%%%%%%%%%%%%%%%%%%%%%%%%%%%%%%%%%%
We consider the Boussinesq model of an incompressible velocity field
${\bf u}({\bf x},t)$ coupled with a temperature (density) field 
$\theta({\bf x},t)$
in the presence of a time dependent acceleration 
${\bf g}(t)=-g(t) \hat{\bf z}$ in the 
vertical direction
\begin{equation}
{\partial {\bf u} \over \partial t}+ {\bf u} \cdot {\bf \nabla} {\bf u} 
= - {\bf \nabla} p + \nu \nabla^2 {\bf u} - \beta {\bf g} \theta
\label{eq1}
\end{equation}
\begin{equation}
{\partial \theta \over \partial t}+ {\bf u} \cdot {\bf \nabla} \theta
= \kappa \nabla^2 \theta
\label{eq2}
\end{equation}
where $\nu$ is the kinematic viscosity, $\kappa$ the thermal diffusivity
and $\beta$ the thermal expansion coefficient.
We assume standard initial conditions for RT, with 
${\bf u}({\bf x},t)=0$ and $\theta({\bf x},0)=-(\theta_0/2) \text{sgn}(z)$
where $\theta_0$ is the temperature jump which defines the Atwood 
number $A=\beta \theta_0/2$.

In the case of constant $g$, the usual phenomenology of RT turbulence is
expected. The initial condition is unstable with respect to perturbation
of the interface $z=0$ and, for a single-mode perturbation at wavenumber 
$k$, the linear stability analysis for an inviscid potential flow gives
the growth rate of the amplitude $\lambda(k)=\sqrt{A g k}$, while
high wavenumber are stabilized by viscosity and diffusivity 
\cite{kull1991theory,zhou2017rayleigh}.
After this linear phase, 
the system enters in a nonlinear regime in which a turbulent mixing layer
grows starting from the interface at $z=0$. Within the
mixing layer, potential energy is converted in turbulent kinetic
energy, with large-scale velocity fluctuations increasing dimensionally
as $u_{rms} \simeq A g t$. Consequently, the width of the
mixing layer $h$ grows as $h(t) \simeq A g t^2$ 
\cite{boffetta2017incompressible}.

In this Letter we consider the case of periodic acceleration $g(t)$. 
To be specific,
we study the case of a square wave of amplitude $g_0$ 
in which the sign of $g(t)$ 
is reversed every half period $T/2$ (starting from the positive, unstable
value).
We also studied the different protocol of sinusoidal
function $g(t)=g_0 \cos(2 \pi t/T)$ (not reported here)
to verify the generality of our results. 

We performed extensive numerical simulations of the Boussinesq equations
(\ref{eq1}-\ref{eq2}) by means of fully parallel pseudo-spectral
code in a $3D$ box of dimension
$L_x=L_y=L_z/4$ at resolutions up to 
$512 \times 512 \times 2048$ with uniform grid.
The boundary conditions for velocity are periodic in the lateral directions
and no-slip in the vertical direction. 
%The height $L_z$ is sufficiently large to allows for the development of 
%the mixing layer. 
Time evolution is obtained by a second-order Runge-–Kutta scheme 
with explicit linear part. 
For all runs we fix $Pr=\nu/\kappa=1$ and viscosity is sufficiently large
to resolve the small scales during all the phases of the simulations
\cite{boffetta2010statistics}.
In all runs, the value of the period $T$ is taken sufficiently long to 
allow for the development of the turbulent mixing layer before
inverting the acceleration, i.e. $\lambda(k) T > 1$.
Numerical results are made dimensionless by using 
$L_x$, $\tau=\sqrt{L_x/(Ag)}$ and $\theta_0$
as unit length, time and temperature respectively.

RT instability is seeded by perturbing the initial temperature field
with respect to the unstable step profile. Two different perturbations were implemented.  
%in order to check the independence of the results from initial conditions. 
In the first case, we perturbed the initial condition by adding 
$5 \%$  of white noise to the value of $\theta({\bf x},0)$ 
in a narrow layer around the interface $z=0$. In the second 
set of simulations, we perturb the interface by a superposition of 
$2D$ linear waves of small amplitudes and narrow range of wavenumbers
around $k=30$. The two sets of simulations gave similar results for
long time statistics.
The results reported in this Letter are relative to the
first type of perturbation, and they are averaged 
over $10$ independent realizations of the white noise for each value
of the parameters. 

%%%%%%%%%%%%%%%%%%%%%%
\begin{figure}[h]
\includegraphics[width=\columnwidth]{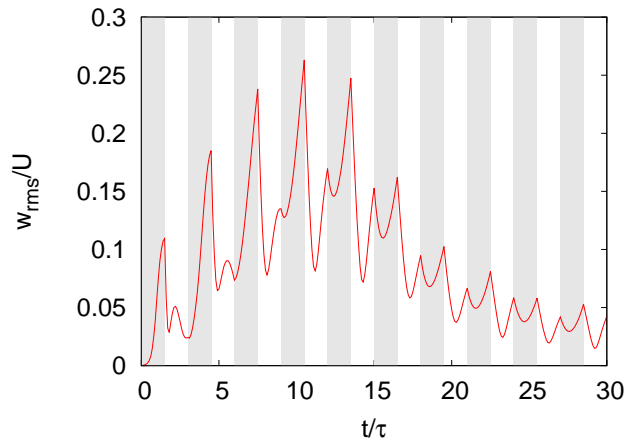}
\caption{Rms of the vertical velocity component in the mixing layer,
made dimensionless with $U=A g \tau$, as a function
of dimensionless time for the case of period $T=3 \tau$. 
Grey (white) regions represent the unstable 
(stable) phases.}
\label{fig2}
\end{figure}
%%%%%%%%%%%%%%%%%%%%%%

Figure~\ref{fig2} shows the evolution of the vertical rms velocity 
for a simulation with period $T=3 \tau$.
At the beginning we observe an average growth of $w_{rms}$
with large oscillations within each period $T$, 
which are due to the acceleration in the unstable phases (grey region) 
and deceleration in the stable phases (white regions). 
Nonetheless, after few periods (the first $4$ periods in this case),
the amplitude of the oscillations reduces and vertical velocity
fluctuations asymptotically decay.
During the unstable phases there 
is on average a growth of the velocity fluctuations, 
initially in agreement with the dimensional linear law. 
After the first periods, the growth of $w_{rms}$ is preceded by
a transient decrease. This is due to the fact that, 
immediately after each acceleration reversal, 
the plumes must invert the direction of their motion before accelerating again. 
 Fig.~\ref{fig2} also shows a secondary peak of vertical velocity
in the stable phases (white regions), 
which is caused by the exchange between kinetic and potential energy
in stably stratified turbulence. 
The period of these secondary oscillations $T_{BV} = 2\pi/N$ 
is determined by the Brunt-V\"ais\"al\"a frequency 
$N=\sqrt{g\beta |\partial \langle \theta\rangle /\partial z |}$ 
where $\langle \theta(z,t) \rangle \equiv 1/(L_x L_y) \int dx \int dy 
\theta({\bf x},t)$ is the mean temperature profile. 
The initial strong gradient of temperature at the interface gives
large values of $N$ and therefore a short secondary period $T_{BV} \ll
T$. 
By approximating the mean temperature gradient  
as $|\partial \langle \theta\rangle /\partial z | \simeq \theta_0/h$, 
we obtain $N \simeq \sqrt{2Ag/h}$. 
Therefore the growth of the mixing layer causes a decrease of the
Brunt-V\"ais\"al\"a frequency. 

%%%%%%%%%%%%%%%%%%%%%%
\begin{figure}[h]
\includegraphics[width=\columnwidth]{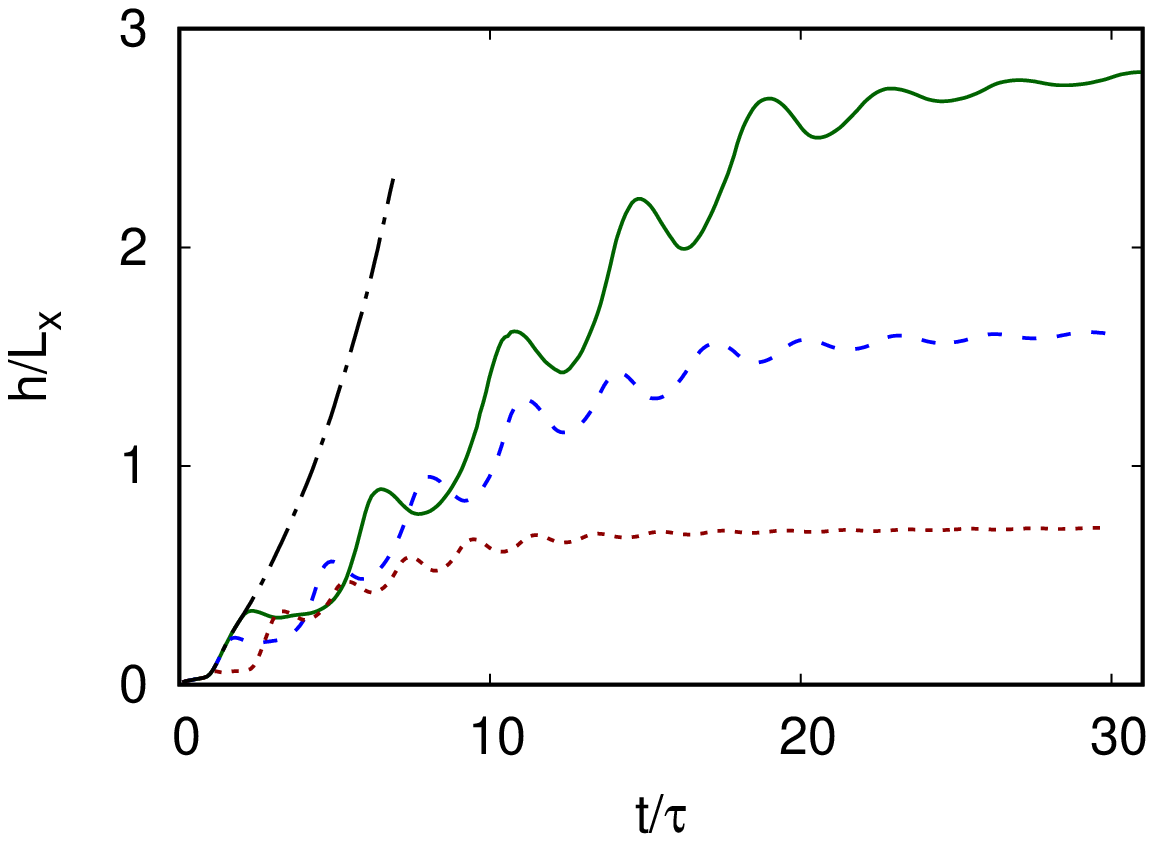}
\includegraphics[width=\columnwidth]{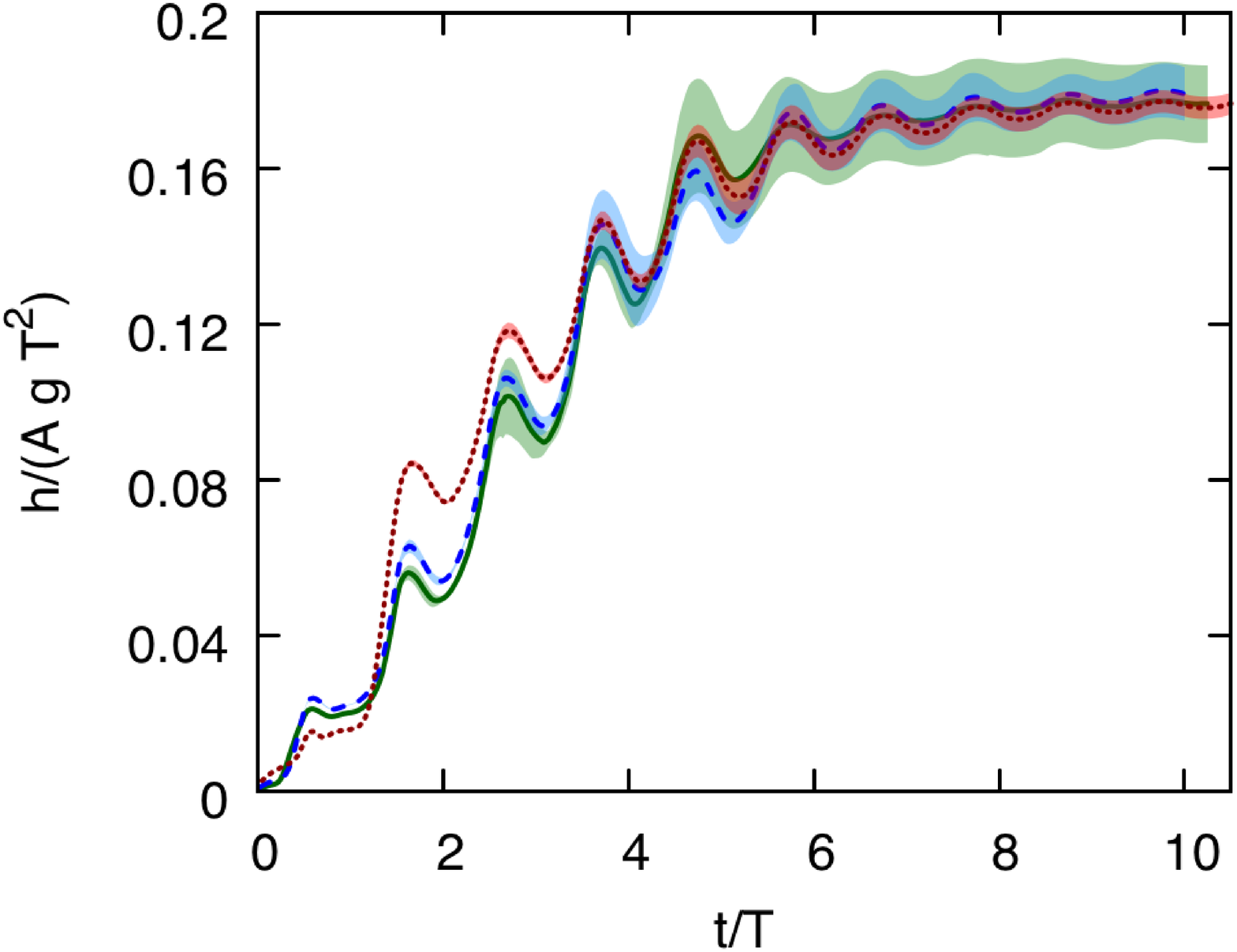}
\caption{(Upper plot) Evolution of the width of the mixing layer for 
$4$ simulations starting from the same set of initial
condition with different periods of the acceleration: 
$T=\infty$ (black dot-dashed 
 line), $T=2 \tau$ (red dotted line), $T=3 \tau$ 
(blue dashed line) and $T=4 \tau$ (green continuous line).
(Lower plot) The same data plotted by rescaling time with
the period $T$ and space with $A g T^2$, with ensemble rms (shadow area).}
\label{fig3}
\end{figure}
%%%%%%%%%%%%%%%%%%%%%%

The long-time decay of velocity fluctuations indicates that asymptotically
the unstable phase is unable to sustain turbulence in the mixing
layer which is eventually arrested. This surprising result is indeed observed
in our simulations, as shown qualitatively in Fig.~\ref{fig1}. 
More quantitatively, in Fig.~\ref{fig3} we show the time evolution of the
width $h(t)$ of the mixing layer, which is defined from the 
mean temperature profiles
%$\langle \theta(z,t) \rangle \equiv 1/(L_x L_y) \int dx \int dy \theta({\bf x},t)$
as the difference between the two heights $z_\pm$ at which 
$\langle \theta(z_\pm,t) \rangle = \pm 0.95 (\theta_0/2)$. 
In all the case investigated, after few oscillations
the width of the mixing layer reaches an asymptotic value $h_\infty$. 
Strictly speaking, a truly asymptotic value cannot be reached because 
of the presence of a diffusive terms in (\ref{eq2}) but, since the 
value of $\kappa$ in our simulations is very small, its effect is
observable only on much longer timescales. 

The period of the accelerations is the only external time-scale in 
the dynamics, therefore one is tempted to rescale time in 
Fig.~\ref{fig3} with $T$ 
and correspondingly the spatial scale with the dimensional expression 
$A g T^2$. With this rescaling we observe an almost perfect collapse 
of the curves $h(t)$ at different periods of oscillation.
The asymptotic width of the mixing layer is found to collpse to
$h_\infty \simeq 0.18 AgT^2$ and the corresponding period of 
the Brunt-V\"ais\"al\"a
oscillations is $T_{BV} \simeq 1.9 T$. 

%%%%%%%%%%%%%%%%%%%%%%
\begin{figure}[h]
\includegraphics[width=\columnwidth]{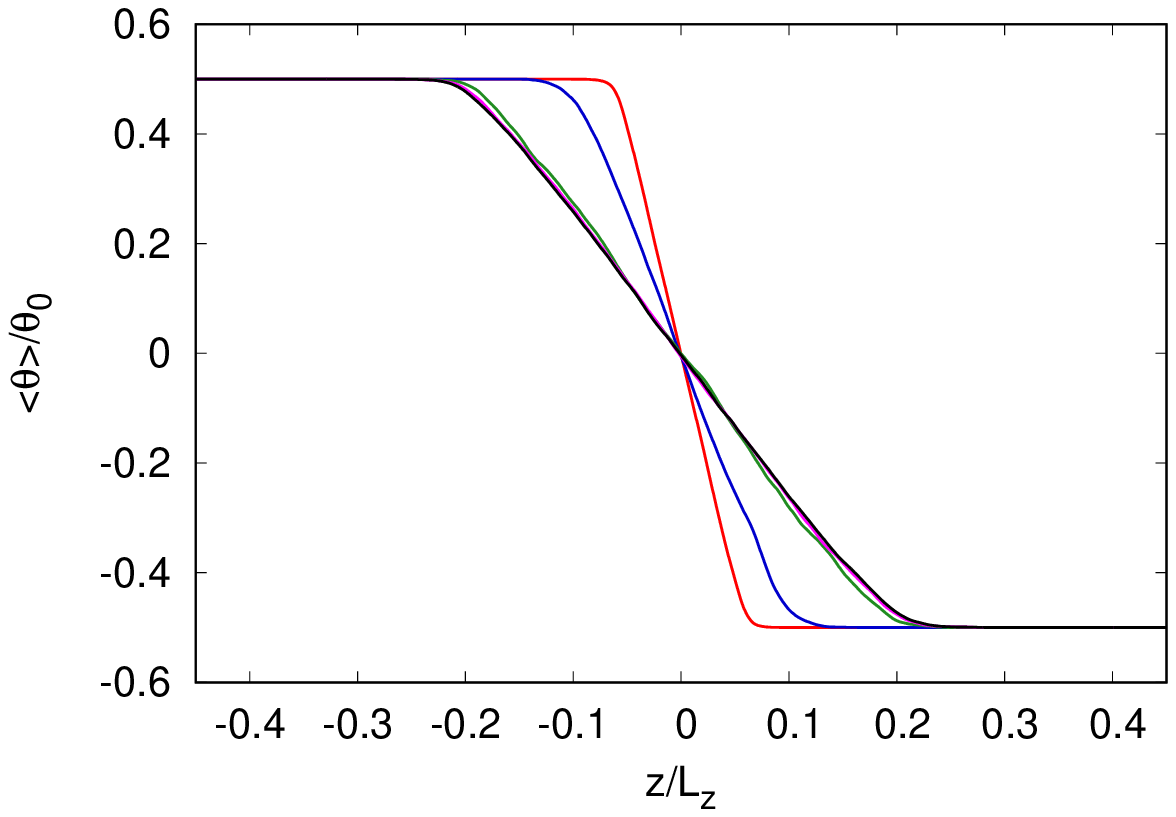}
\includegraphics[width=\columnwidth]{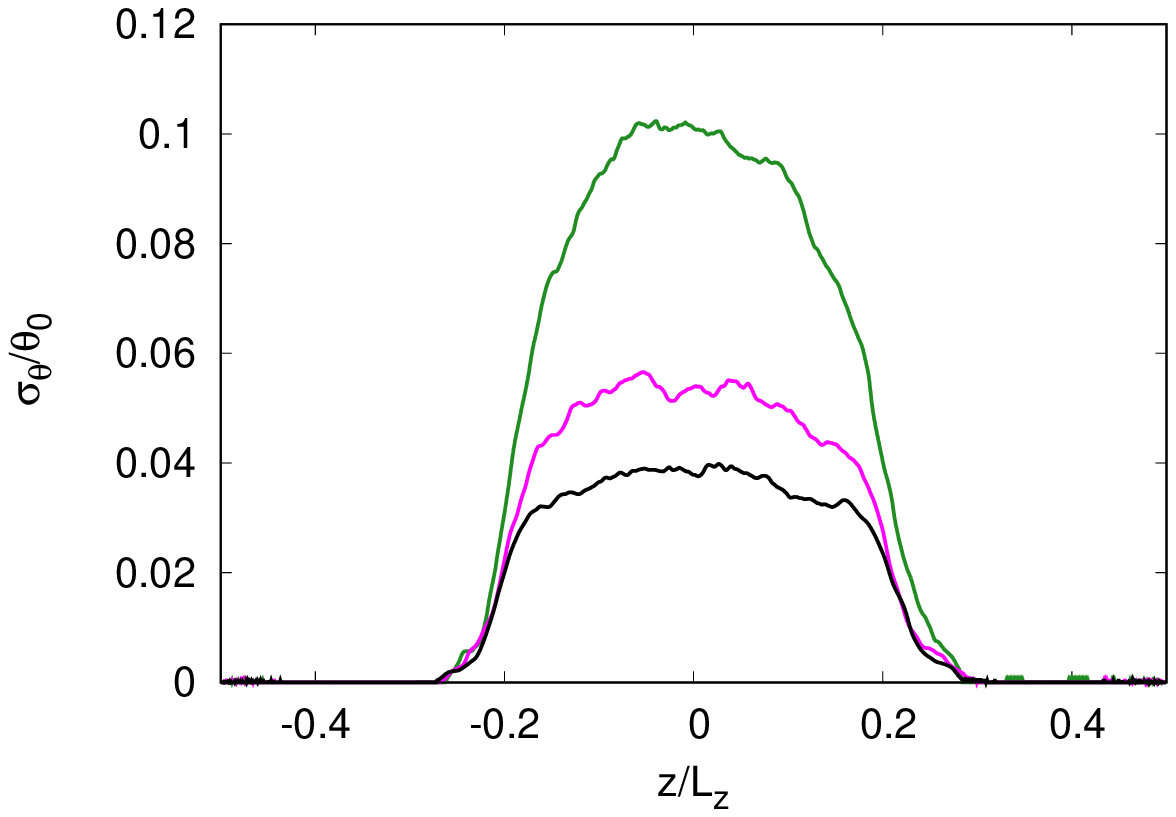}
\caption{(Upper plot) Mean temperature profile 
$\langle \theta \rangle$ for the simulation with period
$T=3 \tau$ at times $t=1.5 T$ (red line), $t=3 T$ (blue line),
$t=6 T$ (green line), $t=8 T$ (magenta line) and $t=10 T$
(black line). 
(Lower plot) Temperature standard deviation $\sigma$ for the simulation with
period $T=3 \tau$ at times $t=6 T$ (green line), 
$t=8 T$ (magenta line) and $t=10 T$ (black line).}
\label{fig4}
\end{figure}
%%%%%%%%%%%%%%%%%%%%%%

Figure~\ref{fig4} shows the mean temperature profiles
$\langle \theta(z,t) \rangle$ and the temperature standard deviation
$\sigma_{\theta}(z,t)=(\langle \theta^2 \rangle - \langle \theta \rangle^2)^{1/2}$
at different times in the evolution of the mixing layer. 
As in usual RT turbulence, the average temperature 
$\langle \theta \rangle$ develops a mean linear profile 
\cite{boffetta2010nonlinear} which evolves in self-similar way
until it is stopped (at $t \ge 8 T$, see Fig.~\ref{fig3}).
At late times the mean temperature profile remains frozen,
while temperature fluctuations, represented by the standard deviation in
Fig.~\ref{fig4}, decay in time, following the decay of velocity fluctuations
shown in Fig.~\ref{fig2}.
All together, these results show that the mechanism which initially 
produces turbulence is unable to sustain the flow for long times. 
Such novel phenomenon of asymptotic relaminarization of turbulence
within the mixing layer in the time-periodic RT system 
is reminiscent of the relaminarization observed in pipe
flows~\cite{eckhardt2007turbulence}.

The physical interpretation of turbulence suppression and of the 
observed rescaling is the following. 
During the stable phase turbulence decays and velocity and temperature
fluctuations are reduced, as shown in Fig.~\ref{fig2} and Fig.~\ref{fig4}.
When the acceleration is switched back to the unstable phase, it takes
some time for the available potential energy to produce new turbulent 
fluctuations and associate kinetic energy, and this time increases
with the width of the mixing layer.
Although the rate of this nonlinear instability cannot be analytically
computed, one can use the results for {\it linear} instability growth
rate, an approximation which is better and better justified with time 
since turbulent fluctuations are decaying.
For the case of a linear temperature gradient of width $h$ which connects
two constant plateaux (a idealized model of the profiles of 
Fig.~\ref{fig4}), linear stability analysis predicts the 
growth rate $\lambda(k)$ of an inviscid perturbation at
wavenumber $k$ as \cite{mikaelian1986approximate}
\begin{equation}
\lambda^2(k) = {A g k \over A + (1-A) h k [1-e^{-h k}]^{-1}}
\label{eq3}
\end{equation}
which recovers the standard result $\lambda^2 = A g k$ in the limit of very
steep gradient $h k \to 0$. In the opposite limit of wide mixing layer, 
which is relevant for the present situation,
$h k \gg 1$ in (\ref{eq3}) gives
\begin{equation}
\lambda^2(k) = {A g \over (1-A) h }
\label{eq4}
\end{equation}
which shows that the growth rate become independent on $k$ and decays
as $1/\sqrt{h}$. Therefore by increasing $h$, 
$1/\lambda$ becomes eventually longer than $T/2$ and 
the perturbation has not sufficient time to grow. From
(\ref{eq4}) this happens at a scale which is proportional to 
$T^2$, in agreement with the phenomenological rescaling used 
in Fig.~\ref{fig3}.
A numerical confirmation of this argument is given by the 
fact that by stopping the periodic acceleration reversal during the 
unstable condition after the asymptotic stage is reached we observe 
the recovery of the quadratic growth of the mixing layer after a time 
longer than $T/2$.

Since RT turbulence is known to produce an efficient exchange of heat
between the two layer at different temperatures, it is interesting
to investigate how this is affected by the acceleration reversal.
We have therefore computed the evolution of the Rayleigh number 
$Ra=A g h^3/(\nu \kappa)$ and of the Nusselt number 
$Nu=\langle w \theta \rangle h /(\kappa \theta_0)$ 
in our numerical simulations. We remind that in standard RT turbulence
both $Ra$ and $Nu$ grow in time (following the growth of $h$ and of
turbulent velocities) and the dependence of $Nu$ on $Ra$ realizes the 
so-called ``ultimate state of thermal convection'' for which
$Nu \simeq Ra^{1/2}$ \cite{kraichnan1962turbulent,ahlers2009heat,
boffetta2009kolmogorov}.

%%%%%%%%%%%%%%%%%%%%%%
\begin{figure}[h]
\includegraphics[width=\columnwidth]{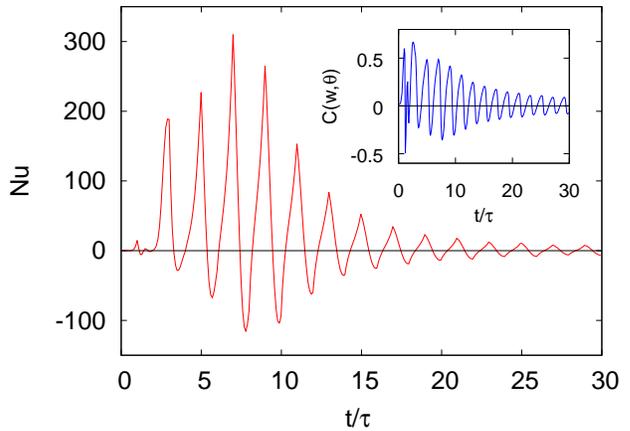}
\caption{Time evolution of the Nusselt number for the case 
$T=2 \tau$. Inset: time evolution of the correlation
$C(w,\theta)=\langle w \theta \rangle/(w_{rms} \theta_{rms})$.}
\label{fig5}
\end{figure}
%%%%%%%%%%%%%%%%%%%%%%

In the present case, since the mixing scale $h$ is arrested,
the Rayleigh number reaches an asymptotic value 
$Ra_{max} \propto T^6$, while the Nusselt number,
shown in Fig.~\ref{fig5}  decays in
time as a consequence of the reduction of velocity fluctuations
(see Fig.~\ref{fig2}). We observe also that $Nu$ is negative
during the stable phase as a consequence of the inversion of the
vertical velocity which becomes anticorrelated with temperature 
fluctuations. 
The decay of $Nu$ is not simply due to the reduction of
vertical velocity fluctuations: indeed also
the correlation $C(w,\theta)=\langle w \theta \rangle/(w_{rms} \theta_{rms})$ 
between the velocity and the temperature fields decreases in time, 
as shown in the inset of Fig.~\ref{fig5}. 
We also observe an increasing symmetry between the stable and the 
unstable phases in the oscillations of $Nu$. 
As a consequence, after few periods of oscillations the total heat
flux over a period is close to zero: the positive flux during the
unstable phases is compensated by an equal and opposite flux in the
stable phase.

%%%%%%%%%%%%%%%%%%%%%%%%%%%%%%%%%
In conclusion, we have studied the phenomenology of Rayleigh-Taylor 
turbulence in the Boussinesq approximation
in presence of a periodic acceleration which alternates phases
of unstable and stable stratification. 
We have found the surprising result that after few periods $T$ of the
acceleration, the turbulent mixing layer reaches a finite extension 
of amplitude proportional to $T^2$. In this asymptotic state
turbulence is found to decay in time, in spite of the persistence
of phases of unstable stratification, and becomes ineffective to 
develop further instabilities.
Extensive numerical simulations show that this result is 
robust in the sense that the presence of an asymptotic layer
is independent on the period $T$ (in the range of period compatible 
with the size of the domain) and also on the specific protocol
of the acceleration (sinusoidal or square wave). 

More in general, our results shed new light on the possibility 
to control, and even suppress, turbulent convection by periodic 
modulation of the acceleration field. This suggests possible 
applications beyond the specific RT configuration.
For example, it is well known that parametric excited Rayleigh-Benard
convection, by thermal or acceleration fast modulation, changes the
onset of the instability for convection 
\cite{niemela1987external,swaminathan2017experimental,ahlers2009heat}. 
Our results show that also the nonlinear, turbulent phase 
could be in principle controlled (and suppressed) by appropriate 
modulation of the external acceleration. 
The possibility to observe this effect
in other systems would be extremely interesting, encouraging 
further numerical or experimental investigations.

%%%%%%%%%%%%%%%%%%%%%%%%%%%%%%%%%%%%%%%%%%%%%%%%%%%%%%%%%%%%%%%%%
\noindent {\sc Acknowledgments.} 
G.B. acknowledges support by the project CSTO162330 
{\it Extreme Events in Turbulent Convection} and from
the {\it Departments of Excellence} grant (MIUR).
M.M. thanks the financial support by the project {\it CRT 2015.2697}.
HPC center CINECA is gratefully acknowledged for computing resources.

%%%%%%%%%%%%%%%%%%%%%%%%%%%%%%%%%%%%%%%%%%%%%%%%%%%%%%%%%%%%%%%%%

%%%%%%%%%%%%%%%%%%%%%%%%%%%%%%%%%%%%%%%%%%%%%%%%%%%%%%%%%%%%%%%%%
\bibliography{biblio}
%%%%%%%%%%%%%%%%%%%%%%%%%%%%%%%%%%%%%%%%%%%%%%%%%%%%%%%%%%%%%%%%%

\end{document}